\documentclass[%
 reprint,
 amsmath,amssymb,
 aps,
]{revtex4-1}

\usepackage{graphicx}% Include figure files
\usepackage{dcolumn}% Align table columns on decimal point
\usepackage{bm}% bold math
\usepackage{dsfont}
\usepackage{color}
\usepackage{ulem}

\begin{document}

\preprint{APS/123-QED}

\title{Comment on scalar glueball mass}
% Force line breaks with

\author{Amir H. Fariborz
	$^{\it \bf a}$~\footnote[1]{Email:
		fariboa@sunyit.edu}}

\author{Renata Jora
	$^{\it \bf b}$~\footnote[2]{Email:
		rjora@theory.nipne.ro}}

\affiliation{$^{\bf \it a}$ Department of Matemathics/Physics, SUNY Polytechnic Institute, Utica, NY 13502, USA}
\affiliation{$^{\bf \it b}$ National Institute of Physics and Nuclear Engineering PO Box MG-6, Bucharest-Magurele, Romania}

\begin{abstract}
We study the interaction of scalar glueball with quark-antiquark and four-quark spinless  meson fields within the framework of a generalized linear sigma model in which the trace anomaly of QCD is exactly realized.   To determine the pure scalar glueball mass ($m_h$), we consider the decoupling limit in which the scalar glueball decouples from the meson fields.  We find the exact relationship $m_h = 2\, h_0$ where $h_0$ is the vacuum expectation value of the scalar glueball field independent of the properties of the framework used.

\end{abstract}
\maketitle

The glueballs are the inherent byproducts of self-interacting gluons in QCD \cite{review}.   Since they can mix with meson fields of the same quantum numbers, determination of their properties becomes naturally challenging.    The scalar glueballs are particularly difficult to study due to the fact that they have the quantum numbers of QCD vacuum and can induce spontaneous chiral symmetry breaking by developing vacuum expectation values.   There have been an extensive investigation of glueballs from different approaches, including lattice QCD \cite{lattice}, QCD sum-rules \cite{QCDSR,Narrison} and  chiral models \cite{chiral}.  

We probe scalar glueballs within the framework of the generalized linear sigma model of \cite{glsm1,glsm2} in which the properties of scalar and pseudoscalar mesons below and above 1 GeV and their underlying quark-antiquark and four-quark mixings have been extensively studied.   In this work we are specifically  interested in properties of the lightest pure scalar glueball, hence, we consider a limit of the generalized linear sigma model in which the glueballs decouple from the quark-antiquark and four-quark states.   The model is constructed in terms of 3$\times$3 matrix
chiral nonet fields:
\begin{equation}
M = S +i\phi, \hskip 2cm
M^\prime = S^\prime +i\phi^\prime,
\label{sandphi1}
\end{equation}
where $M$ and $M'$ transform in the same way under
chiral SU(3) transformations
\begin{eqnarray}
M &\rightarrow& U_L\, M \, U_R^\dagger,\nonumber\\
M' &\rightarrow& U_L\, M' \, U_R^\dagger,
\end{eqnarray}
but transform differently under U(1)$_{\rm A}$
transformation properties
\begin{eqnarray}
M &\rightarrow& e^{2i\nu}\, M,  \nonumber\\
M' &\rightarrow& e^{-4i\nu}\, M'.
\end{eqnarray}
$M$ contains the ``bare" (unmixed)
quark-antiquark scalar nonet $S$ and pseudoscalar nonet $\phi$,  while
$M'$ contains ``bare" two quarks and two antiquarks scalar nonet $S'$ and pseudoscalar nonet $\phi'$.
The model distinguishes  $M$  from $M'$ through the U(1)$_{\rm  A}$ transformation.

The Lagrangian density has the following general structure in terms of chiral nonets $M$ and $M'$ as well as scalar glueball $h$ and pseudoscalar glueball $g$:
\begin{eqnarray}
{\cal L}&=&-\frac{1}{2}{\rm Tr}(\partial^{\mu}M\partial_{\mu}M^{\dagger})-\frac{1}{2}{\rm Tr}(\partial^{\mu}M'\partial_{\mu}{M'}^{\dagger}) 
-V, \nonumber \\
- V&=& f + f_{\rm A} + f_{\rm S} + f_{\rm SB}.
\label{inlgr56}
\end{eqnarray}
Here $f$ is a general function of fields $M$, $M'$, $h$ and $g$ and  is chiral, U(1)$_{\rm A}$ and scale invariant;  $f_{\rm A}$ and $f_{\rm S}$ exactly mock up the axial and trace anomalies and $f_{\rm SB}$ introduces explicit breaking of the chiral symmetry due to quark masses. 
The leading choice of terms
corresponds to eight or fewer underlying quark plus antiquark lines 
at each effective vertex \cite{glsm1,glsm2} and are as follows: 
\begin{eqnarray}
f(M, M', g, h) &=&
-\left(
 u_1 h^2 {\rm Tr}[MM^{\dagger}]
+ u_2{\rm Tr}[MM^{\dagger}MM^{\dagger}]
\right.
 \nonumber \\
&&
\hskip .5cm
  + u_3 h^2 {\rm Tr}[M^{\prime}M^{\prime \dagger}]
\nonumber \\
&&
\hskip .5cm
+ u_4 h (\epsilon_{abc}\epsilon^{def}M^a_dM^b_eM^{\prime c}_f+h.c.)
\nonumber \\
&&
\left.
\hskip .5cm
 + u_5 h^4 + u_6  h^2 g^2  + \cdots \right),
\label{potential7356}
\end{eqnarray}
\begin{equation}
f_{\rm A} =
i{G\over 12} \left[ \gamma_1\ln \left(\frac{\det M}{\det M^{\dagger}}\right)+\gamma_2\ln\left(\frac{{\rm Tr}(MM^{\prime\dagger})}{{\rm Tr}(M^{\prime}M^{\dagger})}\right)\right],
\label{potential73569}
\end{equation}
where $G=h^3 g$ and  $\gamma_1$ and $\gamma_2$  are arbitrary parameters that must satisfy the constraint $\gamma_1+\gamma_2=1$.
\begin{eqnarray}
f_{\rm S} =
-H && \left\{ 
\lambda_1 \ln\left(\frac{H}{\Lambda^4}\right)  \right.
\nonumber \\
 &&  +\lambda_2 \left[\ln\left(\frac{\det M}{\Lambda^3}\right)+\ln\left(\frac{\det M^{\dagger}}{\Lambda^3}\right)\right]
\nonumber \\
&&\left.
+ \lambda_3\left[ \ln\left(\frac{{\rm Tr} MM^{\prime\dagger}}{\Lambda^2}\right)+\ln\left(\frac{{\rm Tr}M' M^{\dagger}}{\Lambda^2}\right)
\right]\right\}.\nonumber \\
\label{scale_anm}
\end{eqnarray}
where $\Lambda$ is the characteristic scale,  $H=h^4$ and $\lambda_1$, $\lambda_2$ and $\lambda_3$  satisfying  the constraint $4\lambda_1+6\lambda_2+4\lambda_3=1$.
The two terms $f_{\rm A}$ and $f_{\rm S}$ ensure that the effective Lagrangian exactly  satisfies the U(1)$_{\rm A}$ and the trace anomalies according to:
\begin{eqnarray}
&&\partial^{\mu}J^5_{\mu}=\frac{g^2}{16\pi^2}N_F\tilde{F}F=G,
\nonumber\\
&&\theta^{\mu}_{\mu}=\partial^{\mu}D_{\mu} = -\frac{\beta(g^2)}{2g}FF=H,
\label{anomaly}
\end{eqnarray}
where $F$ is the SU(3)$_{\rm C}$ field tensor, $\tilde{F}$ is its dual, $N_F$ is the number of flavors, $\beta(g^2)$ is the beta function for the coupling constant, $J^5_{\mu}$ is the axial current and $D_{\mu}$ is the dilatation current.
The symmetry breaking term takes the simple form:
\begin{eqnarray}
f_{\rm SB}=2 {\rm Tr}[AS],
\label{sym3528637}
\end{eqnarray}
where $A={\rm diag}(A_1,A_2,A_3)$ is a  matrix proportional to the three light quark masses.

The isosinglet scalar states $f_1 \cdots f_5$ are a linear combination of the following five components: two- and four-quark building blocks as well as a  glueball component:
  \begin{eqnarray}
(i) \hskip 1cm  f_a&=&\frac{S^1_1+S^2_2}{\sqrt{2}} \hskip .7cm
 \propto \hskip .5cm n{\bar n},
 \nonumber  \\
(ii) \hskip 1cm f_b&=&S^3_3 \hskip 1.6 cm \propto \hskip .5cm s{\bar s},
 \nonumber    \\
(iii) \hskip 1cm f_c&=&  \frac{S'^1_1+S'^2_2}{\sqrt{2}}
 \hskip .5 cm \propto \hskip .5cm ns{\bar n}{\bar s},
 \nonumber   \\
(iv) \hskip 1cm f_d&=& S'^3_3
 \hskip 1.5 cm \propto \hskip .5cm nn{\bar n}{\bar n},
\nonumber   \\
(v) \hskip 1cm h, &&
 \label{f_basis}
 \end{eqnarray}
 where the non-strange ($n$) and strange ($s$) quark content
 for each basis state has been listed at the end of
 each line above.   The VEV of the scalar fields 
 \begin{eqnarray}
\alpha_a &=& \langle S_a^a \rangle\nonumber, \\
\beta_a  &=& \langle {S'}_a^a \rangle\nonumber, \\
 h_0     &=& \langle h \rangle,
 \label{VEVs}
 \end{eqnarray}
with $a=1\cdots 3$ spontaneously break chiral symmetry.

The minimum equations that determine the vacuum of the model are obtained from equations:
\begin{eqnarray}
\left\langle {{\partial V}\over {\partial S_1^1}}\right\rangle_0 &=& 0, \nonumber \\
\left\langle {{\partial V}\over {\partial S_3^3}}\right\rangle_0 &=& 0, \nonumber \\
\left\langle {{\partial V}\over {\partial {S'}_1^1}}\right\rangle_0 &=& 0, \nonumber \\
\left\langle {{\partial V}\over {\partial {S'}_3^3}}\right\rangle_0 &=& 0, \nonumber \\
\left\langle {{\partial V}\over {\partial h}}\right\rangle_0 &=& 0, 
\label{vac_eqs}
\end{eqnarray}
where brackets with subscript zero represent evaluation of each derivative at VEV values of (\ref{VEVs}). 

In order to extract the properties of the scalar glueball,  we consider the decoupling limit in which the scalar glueball decouples from quark mesons:
\begin{eqnarray}
\left\langle {{\partial^2 V}\over {\partial S_1^1\, \partial h}}\right\rangle_0 &=&
4\sqrt{2}u_4[\beta_1\alpha_3+\alpha_1\beta_3] +4\sqrt{2}u_1h_0\alpha_1 \nonumber\\ 
&&+\frac{8\sqrt{2}h_0^3\lambda_2}{\alpha_1}+\frac{8\sqrt{2}h_0^3\lambda_3\beta_1}{2\alpha_1\beta_1+\alpha_3\beta_3}= 0, \nonumber \\
\left\langle {{\partial^2 V}\over {\partial S_3^3\, \partial h}}\right\rangle_0
&=& 8u_4\alpha_1\beta_1 + 4u_1h_0\alpha_3+\frac{8h_0^3\lambda_2}{\alpha_3}  \nonumber \\
&& + \frac{8h_0^3\lambda_3\beta_3}{2\alpha_1\beta_1+\alpha_3\beta_3} =0, \nonumber \\
\left\langle {{\partial^2 V}\over {\partial {S'}_1^1\, \partial h}}\right\rangle_0 &=& 
4\sqrt{2}\alpha_1\alpha_3u_4 + 4\sqrt{2}u_3h_0\beta_1 
\nonumber \\
&&+\frac{8\sqrt{2}h_0^3\lambda_3\alpha_1}{2\alpha_1\beta_1+\alpha_3\beta_3} = 0,  
\nonumber \\
\left\langle {{\partial^2 V}\over {\partial {S'}_3^3\, \partial h}}\right\rangle_0 &=& 
4\alpha_1^2u_4+4u_3h_0\beta_3 \nonumber \\
&&+ \frac{8h_0^3\lambda_3\alpha_3}{2\alpha_1\beta_1+\alpha_3\beta_3}=0.
\label{dclim_eqs}
\end{eqnarray}
The first two minimum equations in (\ref{vac_eqs}) can be used to solve for  $A_1$ and $A_3$ in terms of other parameters. The third and fourth equations in (\ref{vac_eqs}) give:
\begin{eqnarray}
&&4\alpha_1^2\alpha_3\beta_1u_4+2\alpha_3^2\beta_3\alpha_1u_4+2\beta_1^2\alpha_1h_0u_3\nonumber\\
&&+h_0^3\lambda_3\alpha_1+\alpha_3\beta_1\beta_3h_0u_3=0,
\nonumber \\
&&4\alpha_1^3\beta_1u_4+2\alpha_1^2\alpha_3\beta_3u_4+2\alpha_1\beta_1\beta_3h_0u_3\nonumber \\
&& +\alpha_3\beta_3^2h_0u_3+\alpha_3h_0^3\lambda_3=0.
\label{vac_eqs_34}
\end{eqnarray}
Solving this system results in relationships:
\begin{eqnarray}
&&\alpha_3=\alpha_1=\alpha,
\nonumber\\
&&\beta_3=\beta_1=\beta, 
\end{eqnarray}
and consequently $A_1=A_3=A$, and shows that in the decoupling limit the chiral symmetry breaks into its SU(3)$_{\rm V}$ subgroup.   Furthermore, taking the decoupling equations (\ref{dclim_eqs}) into account, we find additional relationships between the model parameters:
\begin{eqnarray}
u_3 &=& -\frac{3\alpha^2u_4}{\beta h_0},\nonumber \\
\lambda_2 &=& -\frac{\alpha^2(h_0u_1+4\beta u_4)}{2h_0^3},\nonumber \\
\lambda_3 &=& \frac{3 \alpha^2\beta u_4}{h_0^3}.
\label{dclim_relations}
\end{eqnarray}

The scalar glueball mass is determined from
\begin{equation}
m_h^2 = \left\langle {{\partial^2 V}\over {\partial^2 h}}\right\rangle_0, 
\end{equation}
where, in the decoupling limit,  becomes
\begin{eqnarray}
m_h^2 &=& 6\,u_1\, \alpha^2 + 6\, u_3 \,\beta^2 + 12\, u_5\, h_0^2 +28
\,{{\it h_0}}^{2}\lambda_1
\nonumber \\
&& + 12\, h_0^2 \left[ \lambda_1\,
\ln  \left( {\frac {h_0^4}{\Lambda^4}} \right) + 2\,\lambda_2\,\ln  \left( {
\frac {\alpha^{3}}{{\Lambda}^{3}}} \right) \right.
\nonumber \\
&&
\left.+2\,\lambda3\,\ln 
\left({\frac {3\,\alpha\,\beta}{{\Lambda}^{2}}} \right)  \right]. 
\label{mh2_raw_SU3}
\end{eqnarray}
The fifth equation in (\ref{vac_eqs}) can be used to solve for $u_5$ in terms of other parameters, which upon substituting this solution  back into (\ref{mh2_raw_SU3}), the glueball mass squared simplifies to:
\begin{eqnarray}
m_h^2 &=& {1\over h_0}
\left(
-36\,{\alpha}^{2}\beta\,u_4 - 12\,u_3\,h_0\,\beta^2 - 12\, u_1\, h_0\,\alpha^2 
\right. \nonumber \\
&&
\left.
+ 16\,h_0^3\lambda_1
\right).
\label{mh_raw_SU3_subs}
\end{eqnarray}
The first two terms cancel out exactly using the first relation in (\ref{dclim_relations}).   Using the other two relations in (\ref{dclim_relations}), simplifies  the third and the fourth terms in (\ref{mh_raw_SU3_subs}):
\begin{eqnarray}
m_h^2&=& - 12\, u_1\,\alpha^2 + 16\,h_0^2\lambda_1 \nonumber \\
     &=&- 12\alpha^2  \left( {{-2h_0^2\, \lambda_2}\over {\alpha^2} } - {{4\beta\, u_4}\over {h_0}} \right) + 16\,h_0^2 \lambda_1\nonumber \\
     &=& 24h_0^2\lambda_2 + 48{{\beta \alpha^2 u_4}\over h_0} + 16\,h_0^2\, \lambda_1 \nonumber \\
     &=& 4h_0^2 \left(4 \lambda_1 + 6 \lambda_2 + 4 \lambda_3\right). 
\end{eqnarray} 
The parenthesis is equal to 1 [see (\ref{scale_anm})] leading to a simple relationship for the pure glueball mass.
\begin{equation}
m_h = 2 h_0. 
\label{mh_h0_relationship}
\end{equation}  

Note that this result which was derived for a specific potential is model independent.  This can be shown by directly examining the trace anomaly for a generic potential $V$ which contains the fields $M$, $M'$, $h$ and $g$.   Requiring that the scale transformation of $V$ exactly gives the second equation of (\ref{anomaly}), yields
\begin{eqnarray}
\left[M_{ij}\frac{\partial V}{\partial M_{ij}}+M_{ij}^{\prime}\frac{\partial V}{\partial M^{\prime}_{ij}}+h.c.\right]
 + \frac{\partial V}{\partial h}h+\frac{\partial V}{\partial g}g
-4V=h^4.
\nonumber\\
\label{trcanom758884}
\end{eqnarray}
Differentiating with respect to the field $h$ and taking the vacuum expectation values results in
\begin{eqnarray}
&&\left[
\left\langle
M_{ij}
\right\rangle_0
\left\langle
\frac{\partial^2 V}{\partial M_{ij}\partial h}
\right\rangle_0
+
\left\langle
M_{ij}^{\prime}
\right\rangle_0
\left\langle
\frac{\partial^2 V}{\partial M^{\prime}_{ij}\partial h}
\right\rangle_0
+h.c.\right]
\nonumber\\
&&
+ 
\left\langle
\frac{\partial^2 V}{\partial h^2}
\right\rangle_0
h_0
+ 
\left\langle
\frac{\partial V}{\partial h} 
\right\rangle_0
+ 
\left\langle
\frac{\partial^2 V}{\partial g\partial h}
\right\rangle_0
\left\langle
g
\right\rangle_0
-
\left\langle
4\frac{\partial V}{\partial h}
\right\rangle_0
\nonumber\\
&&
=  
4 
h_0^3.
\label{trsc7775}
\end{eqnarray}
In the decoupling limit of (\ref{dclim_relations}), the square bracket vanishes.   The first term after the square bracket is the scalar glueball mass squared times $h_0$, terms including  $\left\langle \partial V/ \partial h \right\rangle_0$ vanish because of minimum condition [the last Eq. in (\ref{vac_eqs})] and $\left\langle g \right\rangle_0$ vanishes because of parity, resulting in the same conclusion as in (\ref{mh_h0_relationship}).

\begin{figure}[!htb]
	\centering
	\includegraphics[scale=0.8]{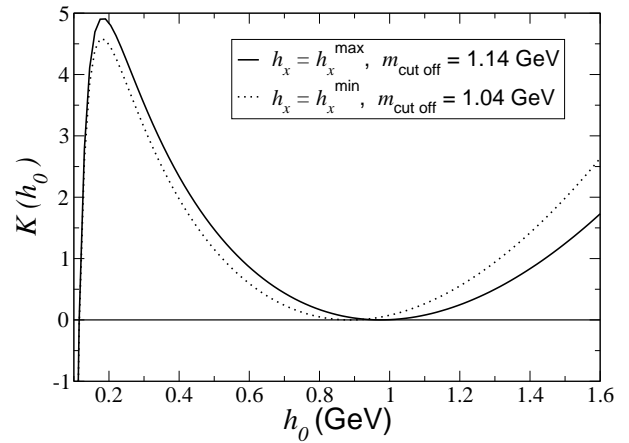}
		\caption{Function $K(h_0)$ [Eq. (\ref{F_h0_hx})] versus $h_0$ with the two values of  $h_x$, its minimum (dotted line) and its maximum (solid line) extracted from QCD sum-rule analysis of \cite{Narrison}.   For each case, the cut off values are adjusted such that $K=0$ occurs at its minimum which is insensitive to extraction of $h_0$.   At the two minima, the values of $h_0$ are 0.887 and 0.974 GeV corresponding to the minimum and maximum values of $h_x$, respectively.  
		}
	\label{F_K_vs_h0}
\end{figure}

Although relationship (\ref{mh_h0_relationship}) can be derived model independently, computation of $h_0$ requires specific modeling of the potential.  
In first  approximation, 
\begin{eqnarray}
h_0^4 \approx h_x^4 = \langle h^4\rangle = \langle H\rangle =\left\langle-\frac{\beta(g)}{2g}G^2\right\rangle=\frac{9}{8\pi}\left\langle \alpha_s G^2\right\rangle, \nonumber \\
\label{h0_hx_aG2}
\end{eqnarray}
where the gluon condensate can be imported from QCD sum-rules or lattice QCD.   However, the relationship between $h_0$ and $h_x$ should be established more carefully.  The partition function for our model reads:
\begin{eqnarray}
Z=\int d h dg dM dM' \exp[i\int d^4 x{\cal L}(h,g,M,M')].
\label{partfunc647766}
\end{eqnarray}
Making a change of variables from  $h$ and $g$ to  $H=h^4$ and $G=h^3g$, with the Jacobian of the transformation
\begin{eqnarray}
J(x-y) = \det\left[H(x)^{-3/2}\delta(x-y)\right],
\label{res5526637}
\end{eqnarray}
the minimum equation for $H$ then becomes:
\begin{eqnarray}
\frac{\partial iS}{\partial H(z)}-{3\over 2}\frac{1}{H(z)}\delta(0)=0.
\label{min7566}
\end{eqnarray}
Here $S$ is the action and the second term comes from the differentiation of the determinant in Eq. (\ref{res5526637}).
Eq. (\ref{min7566}) can be rewritten as:
\begin{eqnarray}
\int d^4y\frac{\partial iS}{\partial h(y)}\frac{\partial h(y)}{\partial H(z)}-{3\over 2}\frac{1}{H(z)}\delta(0)=0,
\label{sec377888}
\end{eqnarray}
which becomes:
\begin{eqnarray}
-i\frac{\partial V(z)}{\partial h(z)}\frac{1}{4H(z)^{3/4}}-{3\over 2}\frac{1}{H(z)}\delta(0)=0.
\label{res663552}
\end{eqnarray}
Note that this time the derivative is not taken in the functional sense. The $\delta$ function in Eq. (\ref{res663552}) needs regularization. Since we are interested in a small range of values around the vacuum condensate we can write:
\begin{eqnarray}
\delta(0)&=&\lim_{x\rightarrow y} \delta(x-y)
\nonumber \\
&=&\int \frac{d^4k}{(2\pi)^4}\exp\left(\frac{\partial^2}{m_c^4}\right)\exp[-ik(x-y)]=
i\frac{m_c^4}{16\pi^2},\nonumber\\
\label{res635529}
\end{eqnarray}
where the integral is computed in the Euclidean space and $m_c$ is an appropriate cut-off scale for a linear sigma model with three flavors.
Then the vacuum equation (\ref{res663552}) becomes:
\begin{eqnarray}
\frac{\partial V}{\partial h}\Bigg|_{h=h_x}\frac{1}{4h_x^3}+\frac{6}{64\pi^2}\frac{m_c^4}{h_x^4}=0.
\label{res6646}
\end{eqnarray}
We multiply the above equation by $4$ and subtract form it the vacuum equation for $h_0$ divided by $h_0^3$ to get:
\begin{eqnarray}
K(h_0) = \frac{\partial V}{\partial h}\Bigg|_{h=h_x}\frac{1}{h_x^3}+\frac{6}{16\pi^2}\frac{m_c^4}{h_x^4}-\frac{\partial V}{\partial h}\Bigg|_{h=h_0}\frac{1}{h_0^3}=0.\nonumber \\
\end{eqnarray}
In the decoupling limit that we considered here, this equation  can be rewritten in terms of the model parameters
\begin{eqnarray}
K(h_0) &=& 
4\lambda_1\ln\left(\frac{h_x^4}{h_0^4}\right) + (6\alpha^2u_1+6\beta^2u_3)\left(\frac{1}{h_x^2}-\frac{1}{h_0^2}\right)
\nonumber\\
&&+12u_4\alpha^2\beta\left(\frac{1}{h_x^3}-\frac{1}{h_0^3}\right)+\frac{6}{16\pi^2}\frac{m_c^4}{h_x^4}=0.
\label{F_h0_hx}
\end{eqnarray}
In this equation, $h_x$ is an input,  computed from Eq. (\ref{h0_hx_aG2}) with the value of the gluon condensate taken from QCD sum-rules or lattice QCD.
(There is a large discrepancy between the values of gluon condensate given by lattice QCD and QCD sum-rules.)
The cut off $m_c$ is adjusted such that $K=0$ occurs at a  stable point (the local minimum of $K$ at which there is no sensitivity to determination of $h_0$).  For example, from QCD sum-rule analysis of \cite{Narrison}:
\begin{eqnarray}
\langle \alpha_s G^2\rangle=(0.070\pm0.013)\,\, {\rm GeV}^4,
\label{secest5566}
\end{eqnarray}
which gives $h_x^{\rm min}=0.057$ GeV and $h_x^{\rm max}=0.083$ GeV, function $K$ vanishes at a local minimum if $m_c$ = 1.04 GeV and 1.14 GeV, respectively, as shown in Fig. \ref{F_K_vs_h0}.    At these local minima $h_0$ is  0.887 GeV and 0.974 GeV, respectively.   These values of $h_0$, when entered into Eq. (\ref{mh_h0_relationship}),  result in a pure scalar glueball mass of 1.77 GeV and 1.95 GeV, overlapping with some of the estimates in the literature \cite{lattice,chiral}.   

In summary,  this work presented a model independent relationship between the mass and the condensate of a pure scalar glueball.   This relationship can be of a key importance in low-energy QCD analyses by bridging frameworks involving quarks and gluons to hadronic models of low-energy QCD.

\section*{Acknowledgments}

A.H.F. gratefully acknowledges  the support of College of Arts and Sciences of SUNY Poly in Spring 2018.

\end{document}